\documentclass{elsart}
\usepackage{epsfig}

\newcommand{\cbcex}{\langle\overline{\chi}\chi\rangle}
\newcommand{\chupiv}{$\chi U\phi_4$}           
\newcommand{\Tr}{\mathop{\rm Tr}\nolimits}

\begin{document}
\begin{frontmatter}
  \title{\hfill {\small J\"ulich, HLRZ 26/97\\
      \hfill Bielefeld, ZIF-MS-10/97\\
      \hfill Cambridge, MIT-CTP-2656\\} 
    \vspace{1cm}
    Chiral transition and monopole  percolation in\\  lattice scalar
  QED with quenched fermions}
\author {Wolfgang~Franzki} 
\address{Institut f\"ur Theoretische Physik E, RWTH Aachen, 
D-52056 Aachen, Germany} 
\author {John~B.~Kogut\thanksref{sabbat}}
\thanks[sabbat]{On sabbatic leave from the University of
  Illinois at Urbana--Champaign}
\address{Center for
    Theoretical Physics, Massachusetts Institute of Technology,\\
    77 Massachusetts Avenue, Cambridge, MA 02139-4307}
\author{Maria--Paola~Lombardo}
\address{Zentrum f\"ur interdisziplin\"are Forschung,\\
  Universit\"at Bielefeld,  D-33615 Bielefeld, Germany}
\date{7 July 1997}
%\maketitle
\begin{abstract}%
  We study the interplay between topological observables and chiral and Higgs
  transitions in lattice scalar QED with quenched fermions. Emphasis is put on
  the chiral transition line and magnetic monopole percolation at strong gauge
  coupling.  We confirm that at infinite gauge coupling the chiral transition
  is described by mean field exponents.  We find a rich and complicated
  behaviour at the endpoint of the Higgs transition line which hampers a
  satisfactory analysis of the chiral transition. We study in detail an
  intermediate coupling, where the data are consistent both with a trivial
  chiral transition clearly separated from monopole percolation and with a
  chiral transition coincident with monopole percolation, and characterized by
  the same critical exponent $\nu \simeq 0.65$.  We discuss the relevance (or
  lack thereof) of these quenched results to our understanding of the \chupiv\ 
  model.  We comment on the interplay of magnetic monopoles and fermion
  dynamics in more general contexts.
\end{abstract}
\end{frontmatter}

\section{Introduction}

Lattice
gauge models with charged scalars and fermions have been studied 
both analytically and numerically for some time
\cite{DeJe92,Shr92}.  Their phase diagrams display a rich structure with chiral
and Higgs transitions, as well as monopole condensation and percolation, and
other topological excitations.  

In this paper we study the interplay of chiral symmetry breaking and monopole
percolation in scalar QED with quenched fermions.  Abelian models can be
rewritten by a duality transformation as theories of topological excitations,
which help in determining their phase structure and critical behaviour
\cite{EiSa78EiSa79}. A classic example is the confinement transition of the
pure compact $U(1)$ lattice model, where monopole condensation was identified
long ago as the underlying physical mechanism \cite{BaMy77}.  It is then
interesting to investigate whether magnetic monopoles affect fermionic
theories, and, in particular, the physics of chiral symmetry breaking and
restoration. The work of \cite{KoWa96x} provides a theoretical framework for
the study of the interplay of monopole and fermion dynamics, and a numerical
strategy for its quantitative verification.  Unfortunately, it has not been
developed into a predictive theory, complete with equations coupling monopole
and fermion observables and their interrelated scaling laws. Rather, it is
just a `physical picture' which could even be internally inconsistent.  The
speculative scenario is interesting and will be summarized here because
we will find some support for it in our simulation data (although more
conventional scenarios can accomodate the data as well).

Consider massless, pointlike fermions propagating in an ensemble of $U(1)$
monopoles. Suppose that the fermions experience only the $U(1)$ gauge
interaction. The fermion--monopole interaction is purely topological in origin
because its strength is given by the product of the electric charge $e$ and
the magnetic charge $g$ which are related by the Dirac quantization condition,
$eg=2\pi$.  This condition is renormalization group invariant, guarantees that
the fermion--monopole interaction is always `strong' and survives conventional
perturbative screening, and lies at the heart of electric-magnetic duality.
Suppose that the monopole network in the vacuum state of such an abelian
$U(1)$ model experiences a second order phase transition at a certain coupling
$e_M$ where the monopole loops grow without bound. Let $\nu$ denote the
correlation length exponent for the transition.  Suppose also that ordinary
photon exchange is not strong enough to cause a chiral transition near $e_M$.
Then, since the fermion's chirality is not a conserved property in the
presence of the pointlike monopole, the monopole transition could induce a
chiral transition and the two point correlation function of $\bar\psi\psi$
should inherit the critical exponent $\nu$ from the monopole--loop transition.
%  (The physical picture for the generation of a nonzero chiral
%  condensate and spatial correlations within it is similar to that of the
%  development of spin order and correlation length scaling in the Ising model.
%  Imagine an Ising system near its critical point $T_c$ so there are large
%  domains of 'up' and 'down' spins with domain walls between them. If two
%  spins, one at position $x$ and the other at $y$, are within one domain, they
%  are strongly correlated. If $x$ and $y$ are in different domains, the
%  spin-spin correlation is diminished.  In this way the order in the spin
%  variable is sensitive to the growth of large domains and the correlation
%  length index $\nu$ controls the fall--off of the spin--spin correlation
%  function.)} 
Past measurements of monopole
percolation have determined $\nu$ with good accuracy, $\nu \simeq 0.67$, which
is near the correlation length exponent of ordinary bond percolation in four
dimensions.  In any case, since $\nu$ is distinct from the mean field value of
$1/2$, this mechanism has been cited as a potential origin of interacting
fermion field theories.

This mechanism has been explored so far in non-compact models
\cite{KoWa96x,KoKo92,HaKo94,KoKo93b,GoHo92x}, where, however, monopole
solutions cannot be found analytically, monopoles are (naively) expected to
decouple in the continuum limit, the Dirac string costs action and static
monopoles do not generate long--range Coulomb fields. The ideal testing ground
for the scenario of \cite{KoWa96x} is a simple model with compact fields, a
second order chiral transition and monopole percolation.  When we begun this
study the $U(1)$ chiral transition was thought to be first order\footnote{The
  recent results on pure compact QED \cite{JeLa96a,JeLa96b,CoFr97b} indicate a
  second order transition and suggest that compact QED could be a very
  interesting testing ground for the scenario discussed in this paper,
  possibly also for the full theory, if negative $\gamma$ is chosen
  \cite{CoFr97c} (extended Wilson action with $-\gamma\sum_P \cos(2\Theta_p)$
  term).}.  These considerations lead us to consider a
fermion--scalar--compact $U(1)$ model, where the existence of a second order
chiral transition was well established both analytically and numerically
\cite{LeShr87a,LeShi86d}.  Note that this transition separates confinement
from Higgs phases, while the QED chiral transition separats confining and
Coulomb phases.

From a field theory point of view, this work continues the search for an
interacting field theory which is strongly coupled at short distance
\cite{Ko94x}.  Considerable work along this line has been invested in the
study of the pure scalar--gauge model, with compact and non-compact gauge
fields.  Detailed numerical studies have supported the idea that non-compact
scalar QED is trivial \`a la $\lambda \phi^4$ by finding evidence for the
tell--tale logarithms \cite{BaFo93,BaFo95}.  Analogous results were reported
for the compact model in \cite{AlAz93}.  The simulation results support the
longheld theoretical prejudice that there is complete charge screening and the
renormalized gauge coupling vanishes.  In addition, in non-compact scalar QED
monopole percolation turns out to be unrelated with the Higgs transition
\cite{BaFo95}.  These results rise the question if the chiral nature of the
transition in fermionic models might make the continuum limit non--trivial.

In a particle physics context, fermion--gauge--scalar models have been
considered as possible toy models for the strongly coupled standard model
\cite{AbFa81a,DaKo88b}.  More recently, one such lattice model -- dubbed in
this context \chupiv\ -- has been reconsidered in
\cite{FrJe95a,FrFr95,FrJe96a,Fr97a,FrJe97b,Fr97b} as a new mechanism for
spontaneous symmetry breaking and dynamical mass generation.  Recent results
showed that the chiral transition line flows into the Higgs transition line,
terminating with a tricritical point, whose critical behaviour indicates
nonpertubative renormalizability, consistent with a Yukawa theory.  The long
term aim is to find out whether and how the physics of the \chupiv\ model is
different from that of the usual Higgs--Yukawa mechanism of fermion mass
generation. For this reason, the triviality issue is not really central here:
it might well be that the model provides an alternative to the usual
Higgs--Yukawa mechanism, with higher upper bounds for its validity.

It would be interesting if the results of this paper could be of use to
further sharpen the understanding of the tricritical point of the \chupiv\ 
model. However, this is a quenched study and its relationship with the full
model can only be judged a posteriori: a better understanding of the \chupiv\ 
model, and/or of the validity of the quenched approximation, is not the
primary scope of this work. Our aim here is to illustrate some
  general dynamical mechanisms possibly relevant to continuum physics
by investigating the interplay between chiral phase transition and
monopole percolation.  We choose to do so in the simplest possible situation:
a scalar $U(1)$ gauge model with quenched fermions.

This paper is organized as follows: first we review the action and phase
diagram of the \chupiv\ model.  We continue by discussing the topological
excitations of Abelian models, and the observables relevant for the chiral
transition.  Section \ref{sec:topex} gives an overview of the results for the
topological excitations and show their correlations with other observables
characterizing the phase diagram. These results confirm and extend the
previous work of \cite{RaKr83}.  After this general study, we focus on three
 points along the chiral transition line: in
Section \ref{sec:perc} we present the results for the percolation of magnetic
monopoles, in Section \ref{sec:chipt} those for the chiral transition. We
conclude with a brief summary and discussion.

\section{Action and phase diagram}
The \chupiv\ model is defined by the action \cite{FrJe95a}
\begin{eqnarray*}
  S &=& S_\chi + S_U + S_\phi \\
  S_\chi & = & \frac{1}{2} \sum_x \overline{\chi}_x \sum_{\mu=1}^4 \eta_{x\mu}
  (U_{x,\mu} \chi_{x+\mu} - U^\dagger_{x-\mu,\mu}
  \chi_{x-\mu}) +{am_0} \sum_x \overline{\chi}_x \chi_x\\
  S_U & = & -\beta \sum_P \cos(\Theta_P) \\
  S_\phi & = & - {\kappa} \sum_x \sum_{\mu=1}^4 (\phi^\dagger_x U_{x,\mu}
  \phi_{x+\mu} + h.c.).
\end{eqnarray*}
$\chi_x$ are the Kogut-Susskind fermion fields, $\phi_x$ is a scalar field with
frozen length $|\phi_x|=1$, and $U_{x,\mu}$ represents the compact link
variable. $U_{x,\mu\nu}$ is the plaquette product of the link variables
$U_{x,\mu}$ and $\eta_{\mu x}=(-1)^{x_1+\cdots+x_{\mu-1}}$.

\begin{figure}
  \begin{center}
    {\epsfig{file=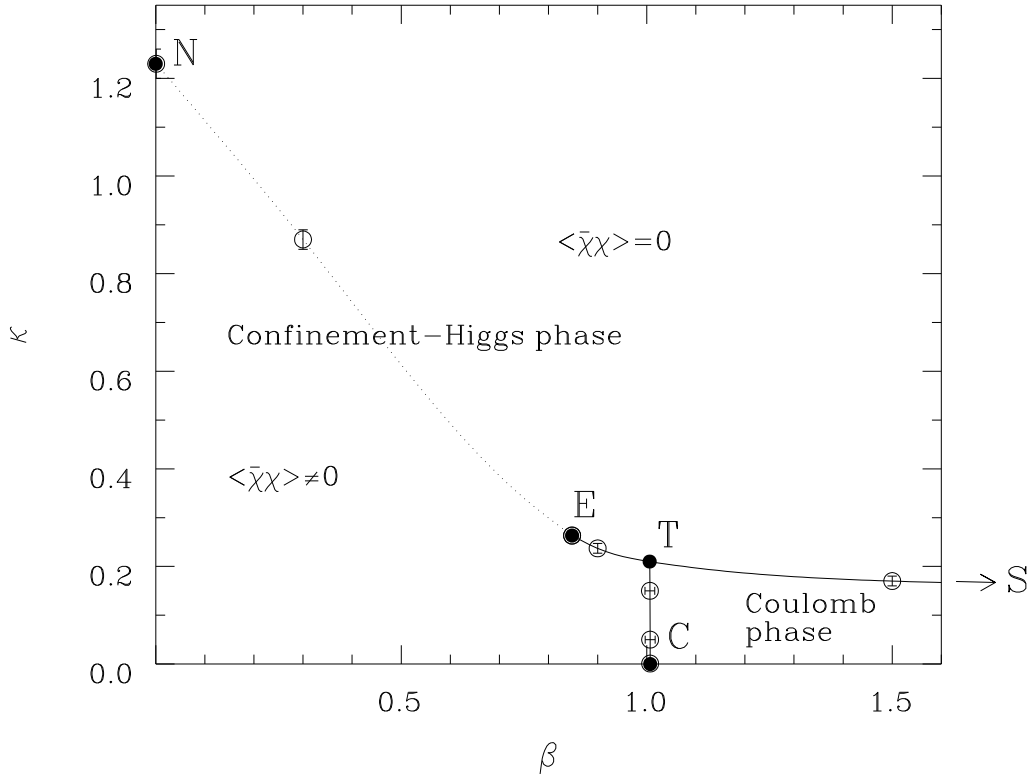,angle=90,width=\hsize,bbllx=65,bblly=150,bburx=540,bbury=780}}%
    \caption[xxx]{%
      The phase diagram of the Higgs model (\chupiv\ model for $am_0=\infty$
      in the update).  The chiral transition of the quenched \chupiv\ model is
      marked dotted.  The points simulated in this work are those marked by
      open circles and the points N and E.  The labels are the same as in
      \cite{FrJe95a,FrFr95}.}
    \label{fig:pd4q2}
  \end{center}
\end{figure}

The phase diagram of the quenched approximation is shown in
figure~\ref{fig:pd4q2}.  It has two phases, the Coulomb phase and the
confinement-Higgs phase. The second is separated into two regions with
different realizations of chiral symmetry,  which are seperate phases
in the chiral limit of the full model. The region with small $\kappa$ and
$\beta$ is chirally asymmetric, the one with large $\beta$ and $\kappa$ is
symmetric.  These two regions are sometimes also called the confinement and
Higgs regions, respectively, because of the limiting cases $\kappa=0$ and
$\beta=\infty$.%

For $\beta=0$ the gauge and scalar fields can be integrated out
exactly, leading to a lattice version of the NJL model \cite{LeShr87a} (point
N). The point E is the critical end point of the Higgs phase transition line
ETC, which is a tricitical point in the chiral limit of the full model. T is
the triple point and C the phase transition from the confinement (at strong
gauge coupling) to the Coulomb phase (at weak gauge coupling) in the model
without the scalar field. S is the critical point of the spin model
($\beta=\infty$).

The status of the current
investigation of the full model, including the dependence on the bare mass,
can be found in \cite{FrFr95,FrJe96a,Fr97a,FrJe97b,Fr97b}.  Even though past
lattice studies in the full model concentrated mostly on the tricritical point
E, the scaling behavior along the chiral phase transition NE except E is
believed to be NJL--like with reasonable accuracy \cite{FrFr95}.  At E the
scaling behavior is different, e.\,g.\ the exponents seem to be nonclassical,
which makes the point particularly interesting.

In this paper we study the quenched model, which corresponds to the limit
$am_0=\infty$ in the update.  The dynamics is the same as in scalar QED, or
Abelian Higgs model, and $S_\chi$ is only considered for the measurement part.
The qualitative features of the phase diagram are known to be the same as in
the full model.  The endpoint of the Higgs transition line E is known with
great accuracy \cite{AlAz93}.  We expect that also in the quenched model the
point N is described by mean field exponents, as observed in \cite{Ho89},
probably with logarithmic corrections.  The topological excitations of the
model have been considered in \cite{RaKr83}.  In our study we will
characterize more fully the phase diagram using topological excitations and we
will study in quantitative detail the chiral transition line extending from
the point N to the endpoint E, and its interplay with the monopole percolation.

\section{Operators}

In Abelian models with compact fields the periodic nature of the interactions
makes it possible to rewrite the Action in terms of its topological
excitations: magnetic monopole loops in pure compact QED, and closed vortex
surfaces for pure scalar matter.  For scalar--gauge systems there are strings
of magnetic flux, i.\,e.\ lattice monopoles, closed vortex surfaces and open
surfaces bounded by monopole loops.  These topological excitations,
investigated in the continuum in \cite{NiOl73}, are amenable to numerical
studies on the lattice \cite{RaKr83,DeTo80,HaWe89}.  We review the main
definitions below.

The study of the chiral transition will follow 
\cite{DaKo91,KoKo93a,LoKo94}.
The relevant operators are reviewed in the last subsection.

\subsection{Monopoles}
For the measurement of the monopole loop density we introduce 
the flux variables \cite {DeTo80}
\begin{equation}
  \theta_{\rho\sigma}(x) = \theta_{x,\rho} + \theta_{x+\rho,\sigma} -
  \theta_{x+\sigma,\rho} - \theta_{x,\sigma} \in(-4\pi,4\pi]
\end{equation}
where $\theta_{x,\rho}\in(-\pi,\pi]$ is the phase angle
of the link variables $U_{x,\rho}$
and the physical flux
\begin{equation}
  \bar\theta_{\rho\sigma}(x) = \theta_{\rho\sigma}(x) +2\pi n_{\rho\sigma}(x)
  \in(-\pi,\pi]\;,
\end{equation}
where $n_{\rho\sigma}(x)$ is an integer. If $n_{\rho\sigma}(x)\neq 0$ a Dirac
sheet goes through the plaquette. The balance of the flux entering and leaving
a 3-cube defines (on the dual lattice) a monopole current
\begin{equation}
  2\pi m_\mu(x) = \varepsilon_{\mu\nu\rho\sigma} \nabla_\nu
  \bar\theta_{\rho\sigma}(x)\;.
\end{equation}
The operator $\nabla_\nu$ labels lattice differentiation. The vector
$m_\mu(x)$ is the total flux out of the cube $c$ at the dual site $x$
 in direction $\mu$:
\begin{equation}
  2\pi m_\mu(x) = \sum_{p\in\partial c}\bar\theta_p = \sum_{p\in\partial
    c}(\theta_p+2\pi n_p) = 2\pi\sum_{p\in\partial c}n_p\;.
\end{equation}
The loops of the monopoles (on the dual lattice) are closed,
because of current conservation
\begin{equation}
  \nabla_\mu m_\mu(x) = 0\;.
\end{equation}
They are gauge invariant objects, whereas the Dirac sheets, apart
from their boundaries, can be distorted. 

From this we calculate the monopole loop density, in short 
monopole density, as
\begin{equation}
  \rho_m = \frac{1}{V} \left\langle \sum_{x,\mu}|m_\mu(x)| \right\rangle\;.
\end{equation}

Monopole percolation is detected using an order parameter borrowed
from standard percolation models \cite{St79}. 
A connected cluster of monopoles
is introduced \cite{HaWe89}: one counts the number of dual sites $n$
which are connected with each other by monopole line elements.
$n$ is the size of the cluster.
Note that this construction ignores the vector structure of 
the monopole currents. 

The density of the occupied bonds reads
\begin{equation}
  p_m = \frac{n_{\rm tot}}{4V} = \frac{1}{4V}\sum_{n=4}^{n_{\rm max}}
  g_n n\;,
\end{equation}
where  $g_n$ is the number of clusters with size $n$,
$n_{\rm max}$ is the size of the largest cluster, and $n_{\rm tot}$
is the total number of connected sites.

An order parameter for percolation is
\begin{equation}
  M_{\rm perc} = \frac{n_{\rm max}}{n_{\rm tot}}\;.
\end{equation}
Its associated susceptibility is
\begin{equation}
\chi_M=V\left(\langle M_{\rm perc}^2\rangle - \langle M_{\rm perc}\rangle^2\right).
\end{equation}
We have also considered the same susceptibility as in \cite{HaWe89}
\begin{equation}
  \chi_{\rm perc} = \left\langle \left[ \sum_{n=4}^{n_{\rm max}} g_n n^2
  - n_{\rm max}^2\right] /n_{\rm tot} \right\rangle\;.
\end{equation}

\subsection{Vortex sheets}
For the measurement of the vortex sheets we introduce
\cite{EiSa78EiSa79,RaKr83}
\begin{equation}
  \bar\vartheta_\nu(x) = \varphi_x - \theta_{x,\nu} - \varphi_{x+\nu} +
  a_\nu(x) \in(-\pi,\pi] \;,
\end{equation}
where $\varphi_x\in(-\pi,\pi]$ is the phase angle of $\phi_x$ and
$a_\nu(x)$ is a integer.  The local vortex sheet is now
\begin{equation}
  V_{\rho\sigma}(x) = \varepsilon_{\rho\sigma\mu\nu}\left( \nabla_\mu a_\nu(x)
    + {\textstyle \frac{1}{2}} n_{\mu\nu}(x) \right)\;.
\end{equation}
From this we calculate the vortex density as
\begin{equation}
  \rho_V = \frac{1}{6V} \left\langle \sum_{x,\rho\sigma} |V_{\rho\sigma}(x)|
  \right\rangle\;.
\end{equation}

\subsection{Chiral condensate}
The chiral condensate is measured in the usual way
\begin{equation}
  \cbcex = \left\langle \Tr M^{-1} \right\rangle
\end{equation}
with a stochastic estimator, where $M$ is the fermion matrix.

We also measured the logarithmic derivative of the chiral condensate
\cite{KoKo93a}:
\begin{equation}
  \label{def_rpi}
  R_\pi = \left.\frac{\partial\ln\cbcex}{\partial\ln
      am_0}\right|_{\beta,\kappa} = \left.
    \frac{am_0}{\cbcex}\frac{\partial\cbcex}{\partial am_0}\right|_{\beta,\kappa}\;.
\end{equation}
This can be expressed as ratio of zero momentum meson propagators $C(p=0)$
(susceptibilities):
\begin{equation}
  \label{rpi_sus}
  \left.\frac{\partial\cbcex}{\partial am_0}\right|_{\beta,\kappa} =
  C_\sigma(p=0)\;,\quad\quad
  \frac{\cbcex}{am_0} = C_\pi(p=0)\;,
\end{equation}
where the second is the Ward identity, which results from the chiral
$U(1)_A$ symmetry of the staggered fermions \cite{KiSh87}. 
This gives
\begin{equation}
  R_\pi = \frac{C_\sigma(p=0)}{C_\pi(p=0)}\;.
\label{eq:rpi}
\end{equation}
We note that in the quenched approximation only the connected part of
$C_\sigma$ need to be considered, since fermion loops are neglected
\cite{JF}.  We have checked that our measurement of
$R_\pi$ using eq.~(\ref{eq:rpi}) is in good agreement with the logarithmic
derivative of the chiral condensate computed by numerical differentiation of
the $\cbcex$ results.

\section{Measurement of the topological excitations -- Overview}
\label{sec:topex}
\begin{figure}
  \begin{center}
    \psfig{file=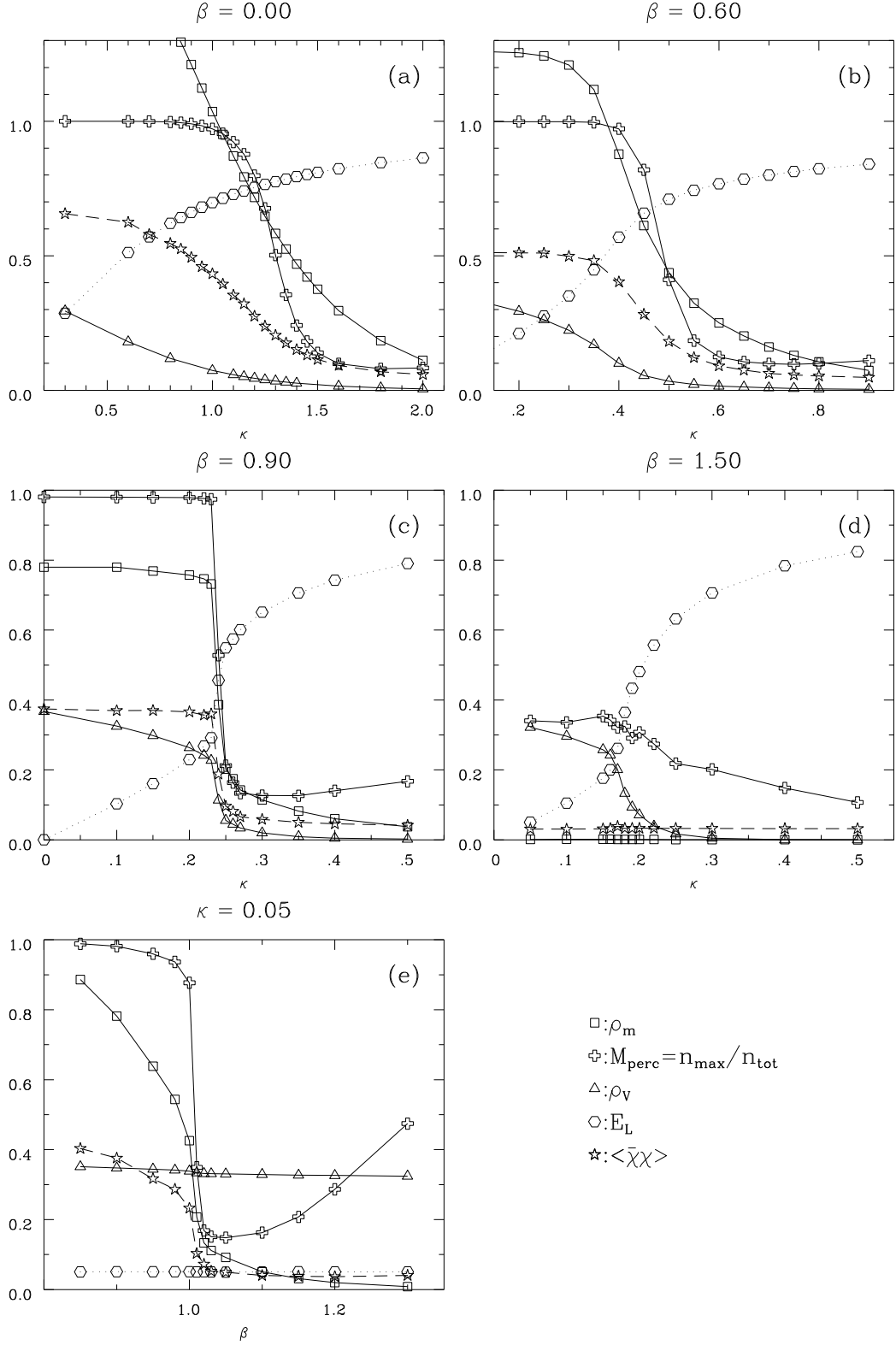,angle=90,height=20.9cm,bbllx=48,bblly=70,bburx=810,bbury=576}%
    \caption{%
      Overview of the topological observables $\rho_m$, $M_{\rm perc}$ and
      $\rho_V$ together with $E_L$ and $\cbcex$ on the $6^4$ lattice at the lines
      NE (a+b), ET (c), TS (d) and CT (e). The errors (not shown) are nearly
      always smaller than the symbols.}
    \label{fig:top}
  \end{center}
\end{figure}

Fig.~\ref{fig:top} shows a overview of the results on the $6^4$ lattice. For
$\beta=0.00$ (a) (NJL line) all observables are very smooth, only the ratio
$M_{\rm perc}$ and the chiral condensate show a somewhat steeper descent.
Monopole and vortex density decrease slowly for increasing $\beta$. More to
the right on the NE line for $\beta=0.60$ (b) a similar behavior is observed
but the transition region is somewhat smaller.%

For $\beta=0.90$ (c) (crosswise to the ET line first order)
a steep descent in all observables at $\kappa=0.24(1)$ can be observed, as
this is typical for first order phase transitions. Obviously the monopoles and
the vortices are sensitive to this phase transition.

At $\beta=1.50$ (d) (crosswise to the TS line first order) a
phase transition from Coulomb to Higgs phase can be observed. At this phase
transition only the vortex density and the link energy are sensitive, which
vary rapidly at $\kappa=0.17$. The other observables stay nearly constant.

The phase transition CT from confinement to Coulomb phase is shown for
$\kappa=0.05$ (e). At the phase transition at $\beta=1.01(1)$ the monopole
density, $\rho_m$ and the chiral condensate steeply decrease.  Link energy and
vortex density are insensitive to this phase transition.

So, we have the following picture. Consider the three regions of the phase
diagram: Coulomb phase, Higgs region and confinement region. In the
confinement region the monopole and vortex density are large and in the Higgs
region both are small. In the Coulomb phase the vortex density is large and
the monopole density small. At the first order phase transtions
the observables which are large in one phase and small in the other
show a steep decrease, whereas the others are insensitive.

The order parameter of percolation $M_{\rm perc}$ follows (essentially) the
monopole density but with more rapid changes.  This could be understood in
analogy to bond percolation.  If the monopole density is high, percolating
clusters are very likely, whereas for small monopole densities they are very
unlikely.  The first percolating clusters occur at some intermediate coupling.
For large $\beta$, along the line of first order phase transitions, this seems
to coincide with the phase transition of the monopole density.  Such interplay
of monopole condensation and percolation has been investigated before in
pure compact QED in \cite{BaFo94b}.

At the line NE, which is second order in the full theory, 
the situation is less obvious.

\begin{figure}%
  \begin{center}%
    \epsfig{file=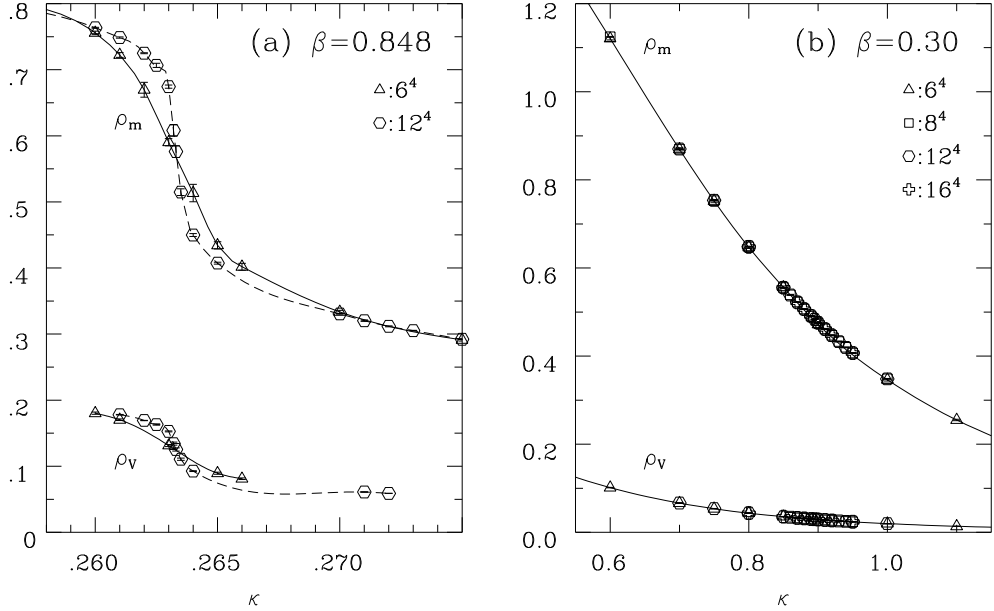,angle=90,width=\hsize,bbllx=65,bblly=25,bburx=540,bbury=810}%
    \caption{%
      Monopole density $\rho_m$ and vortex density $\rho_V$ for (a)
      $\beta=0.848\simeq\beta_{\rm E}$ and (b) $\beta=0.30$.}%
    \label{fig:rhomvor}%
  \end{center}%
\end{figure}

At the endpoint E of the Higgs phase transition, $\beta=0.848$, $\rho_m$ and
$\rho_V$ show a steep descent, which gets increasingly steeper for increasing
lattice size (Fig.~\ref{fig:rhomvor}a).  The derivative of these observables
show an increasing maximum. We have not done a scaling investigation but the
data suggest a divergence.%

For $\beta=0.30$ (Fig.~\ref{fig:rhomvor}b), still along the line NE, both
densities show no maximum in the derivative.  Moreover, no volume dependence
was observed.  Obviously they do not get critical in the region where the
chiral condensate has its largest variation.

For $\beta=0.00$ the picture is very similar  to that at
$\beta=0.30$.

The three $\beta$ values 0.00, 0.30, and 0.848 have been selected for the
systematic investigation which we describe in the rest of this paper.

\section{Magnetic Monopoles  in the vicinity of the NE line}
\label{sec:perc}

In this Section we measure the critical line for the percolation of the
magnetic monopole clusters, and its critical exponents.  An accurate
measurement of the critical line is obtained thanks to the combined use of
reweighting techniques, and the two definitions of monopole percolation
$\chi_{\rm perc}$ and $\chi_M$.

\begin{figure}%
  \begin{center}%
    {\epsfig{file=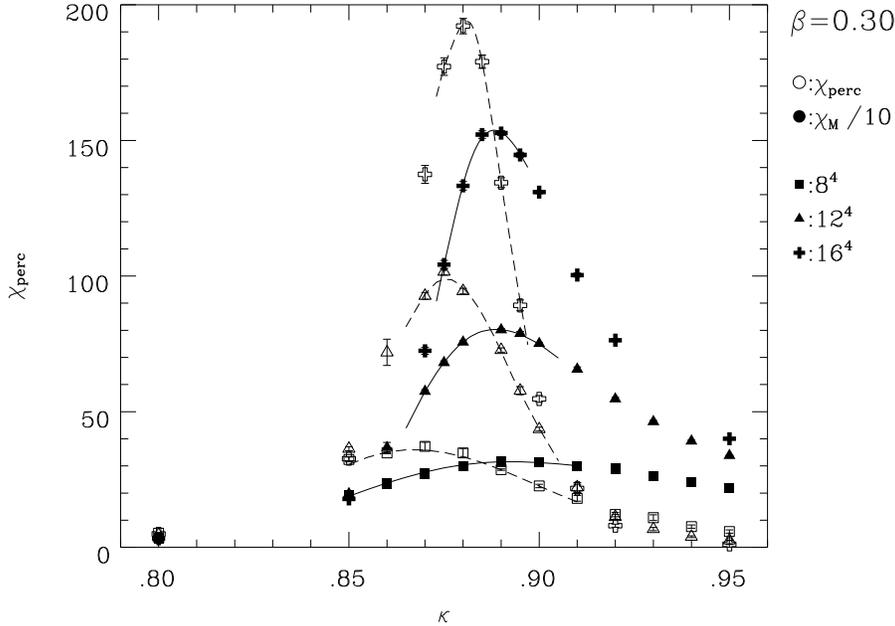,angle=90,width=\hsize,bbllx=65,bblly=15,bburx=540,bbury=780}}%
    \caption{%
      Percolation susceptibilities $\chi_{\rm perc}$ (solid) and 
      $\chi_M/10$
      (open symbols) for $\beta=0.30$. The
      curves are interpolations made with the use of reweighting techniques.}
    \label{fig:twodef}
  \end{center}
\end{figure}

The quality of our measurements is demonstrated in Fig.~\ref{fig:twodef} where
we contrast $\chi_{\rm perc}$ and $\chi_M$ at $\beta=0.30$ on lattices of
different size.  Both susceptibilities show a clear maximum for all $\beta$'s
in the vicinity of the line NE.  The peaks are very
close, and the volume corrections have different signs.%

\begin{figure}
  \begin{center}
    {\epsfig{file=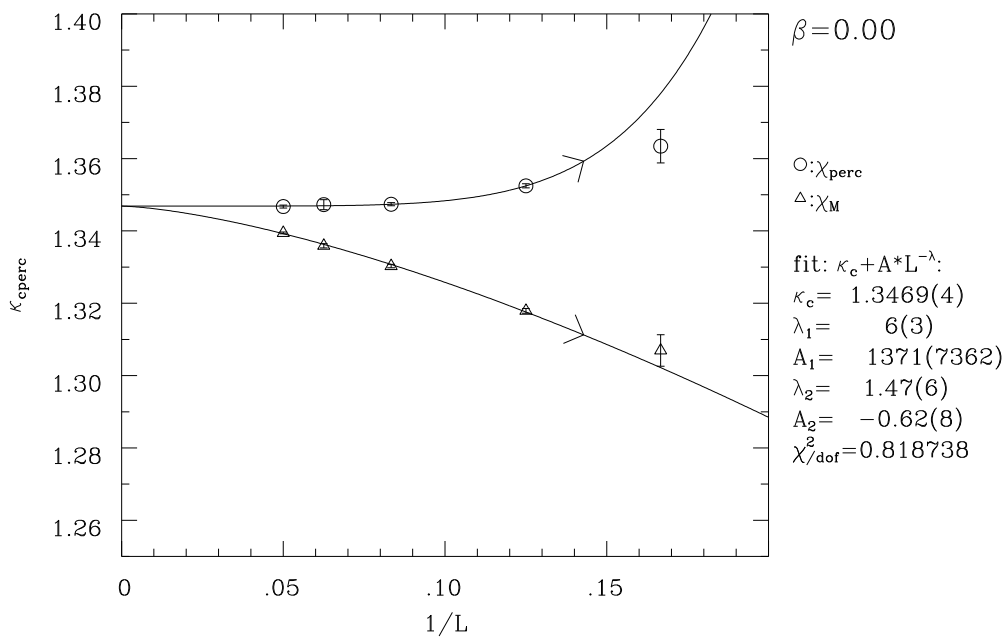,angle=90,width=\hsize,bbllx=65,bblly=15,bburx=540,bbury=780}}%
    \caption{%
      Position of the peak of the percolation susceptibility $\chi_{\rm perc}$
      (circles) and $\chi_M$ (triangles) for $\beta=0.00$.}
    \label{fig:twodef0}
  \end{center}
\end{figure}

We have determined the precise position of the maximum of $\chi_{\rm perc}$
and $\chi_M$ with the multi-histogram method. The numerical results for
$\chi_{\rm perc}$ and $\chi_M$, together with other magnetic observables, are
summarized in table~\ref{tab:suspeak}.  In Fig.~\ref{fig:twodef0} we show the
position of the peak of the two susceptibilities as a function of $1/L$ for
$\beta=0.00$. We superimpose a power law fit with different amplitudes and
exponents and a common critical point.  Again we see that, by increasing
volume, $\chi_{\rm perc}$ approaches the critical point from above, $\chi_M$
from below.  They provide upper and lower bounds, hence an accurate
determination, of the critical coupling for the percolation of magnetic
monopole clusters.%
\begin{table}
  \caption[xxx]{%
    Position and height of the maximum of the percolation susceptibility for
    different $\beta$ and lattice sizes along the line NE.  The values are
    given for both definitions $\chi_M$ and $\chi_{\rm perc}$.  For the
    infinite volume extrapolation of $\kappa_{\rm max}$ ($\kappa=1.3469$,
    $\kappa=0.8875$, and $\kappa=0.2714$, respectively) also the density
    of the occupied bonds $p_m$ and the monopole density $\rho_m$ is printed.
    For comparison, our best estimate for the chiral transition in the
    infinite volume is given in the second column (see section
    \ref{sec:chipt}).}
  \label{tab:suspeak}
  \vspace{2mm}
  \scriptsize
  \renewcommand{\arraystretch}{1.1}
  %% \begin{tabular}{|lr|r|lD{.}{.}{5}|l l }
  \begin{tabular}{@{}l r | l | l r | l r |l l@{}}
    %% \hline
    \multicolumn{1}{c}{$\beta$} &
    \multicolumn{1}{c|}{$\kappa_{{\rm chiral}, c}$} & lattice & 
    \multicolumn{1}{c}{$\kappa_{\rm max}(L)$} &
    \multicolumn{1}{c|}{$\chi_M (L)$} &
    \multicolumn{1}{c}{$\kappa_{\rm max}(L)$} &
    \multicolumn{1}{c|}{$\chi_{\rm perc}(L)$} &
    \multicolumn{1}{c}{$p_m$} & \multicolumn{1}{c}{$\rho_m$}
    \\\hline
    & & $ 6^4$ & 1.307(4) & 156.(3)&   1.363(5) & 16.11(5) & 0.1132(1) &  0.5283(6)\\
    & & $ 8^4$ & 1.3179(6)& 342.(2)&  1.3525(6) &  32.00(4) & 0.11323(2) &  0.52852(9)\\ 
    0.00 & 1.23(3) & $12^4$ & 1.3303(5) & 941.(9) &  1.3474(4) &  80.9(4) & 0.11322(1) & 0.52848(5)\\ 
    & & $16^4$ &1.3358(6)& 1865.(13)&   1.347(1) &  154.(7) & 0.11322(1) & 0.52847(7)\\ 
    & & $20^4$ &1.3395(3)& 3054.(31)&  1.3467(4) &  253.(3) & 0.11321(1) & 0.52841(4)\\ \hline
    & & $ 6^4$ &0.861(4) & 165.(3)&   0.903(3) &  15.6(2) & 0.1074(3)  &   0.494(1)\\ 
    & & $ 8^4$ &0.8671(5)& 360.(2)&   0.8926(8) &  31.5(1) & 0.10746(3) &  0.4946(2)\\ 
    \raisebox{1.2ex}[-1.2ex]{0.30} & \raisebox{1.2ex}[-1.2ex]{0.87(2)}  &
        $12^4$ &0.8756(4)&989.(5) &  0.8886(8) &  80.3(4) & 0.10749(2) &  0.49474(9)\\ 
    & & $16^4$ &0.8812(4)&1937.(13)&  0.8882(5) &  153.7(7) & 0.10750(1) & 0.49477(7)\\
    \hline 
    & & $ 8^4$ &0.2678(1)& 574.(4)&  0.2732(1) & 22.35(6) & 0.0727(2) &  0.3180(7)\\ 
    & & $12^4$ & 0.2692(1)&1437.(15)&0.27204(7) &  59.5(3) & 0.07257(5) &  0.3174(2)\\
    \raisebox{1.2ex}[-1.2ex]{0.848} & \raisebox{1.2ex}[-1.2ex]{0.26:0.29} 
      & $16^4$ & 0.2701(1)&2758.(34)&0.27191(8) & 120.(1) & 0.07259(7)  & 0.3175(3)\\
    & & $20^4$ & 0.27065(5) & 4501.(53) & 0.27168(4)& 206.1(8) & 0.7257(2) & 0.3174(1)
  \end{tabular}
\end{table}%

A determination of the critical exponent of comparable accuracy requires a
careful consideration of scaling violating effects, and will be presented
elsewhere. For our present purposes a precision comparable to that
we achieved at the chiral transition would suffice (5--10
\%).

\begin{figure}
  \begin{center}
    \epsfig{file=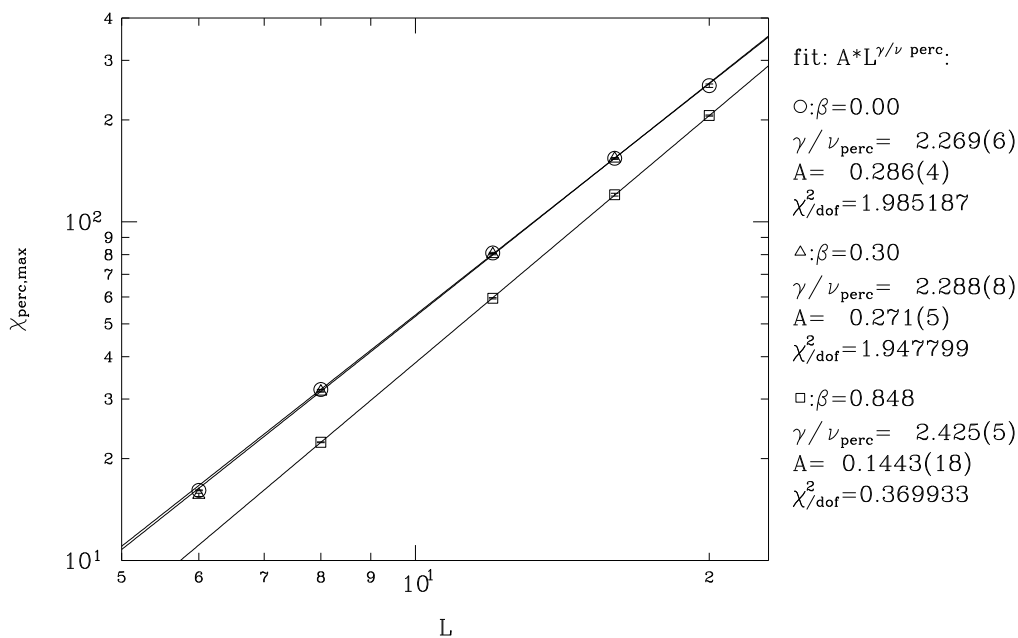,angle=90,width=\hsize,bbllx=65,bblly=15,bburx=540,bbury=780}%
    \caption{%
      Scaling behavior of the maximum of the percolation susceptibility
      $\chi_{\rm perc}$ for $\beta=0.00$, $\beta=0.30$ and $\beta=0.848$. In
      the fit only data with $L\geq 8$ have been considered.}
    \label{fig:suspeak}
  \end{center}
\end{figure}%
Fig.~\ref{fig:suspeak} shows the scaling behavior of the maximum of $\chi_{\rm
  perc}$ with the lattice size.  Note the linear rise with size in this double
logarithmic plot, supporting power law scaling. Only the  data on the
$6^4$ lattice displays  small deviations. In a fit with
the finite size scaling equation
\begin{equation}
  \chi_{{\rm perc},{\rm max}}(L) \propto L^{\gamma_{\rm perc}/\nu_{\rm perc}}
\end{equation}
we measure for $\beta=0.00$ and $\beta=0.30$ the nearly identical value of
$\gamma_{\rm perc}/\nu_{\rm perc}=2.28(2)$. The subscript `perc' is added to
indicate the definition of the exponents at the percolation transition.  For
comparison, the measurements in quenched non-compact \cite{KoKo92} and in full
non-compact QED \cite{KoWa96x} yields quite similar values of $\gamma_{\rm
  perc}/\nu_{\rm perc}=2.24(2)$ and $\gamma_{\rm perc}/\nu_{\rm
  perc}=2.25(3)$, respectively.  For $\beta=0.848$ one obtains a somewhat
larger value of $\gamma_{\rm perc}/\nu_{\rm perc}=2.42(1)$.

The maximum of $\chi_M$ shows larger deviations from the leading finite size
scaling behaviour for small lattices.  If we use only lattices with $L\geq
12$, we get for $\beta=0.00$ and $\beta=0.30$ an exponent
$\gamma_M/\nu_M=2.32(3)$ and for $\beta=0.848$ $\gamma_M/\nu_M=2.23(4)$.  The
subscript $M$ indicates the exponents which have been determined using
$\chi_M$.

The $\gamma/\nu$'s from the two definitions are separated by less than two
standard deviations at $\beta = 0.00$ and $\beta = 0.30$.  At $\beta = .848$
the disagreement increases.  Larger lattices, and/or a careful consideration
of correction to scaling terms should bring the results from the two
definitions closer.  Within this uncertainty all numbers of $\gamma/\nu$ of the
percolation transition are compatible with 2.3(1).

We can contrast these results with those of pure random bond percolation
\cite{GP}.  For pure random bond percolation the critical density is
$p_c=0.161$. In our model the density $p_{m,c}$ at the percolation phase
transition depends on $\beta$, and it is generally lower.  Our result is close
to that of site bond percolation $\gamma/\nu=2.094(2)$ \cite{BaFe96,GP}.
Considering the vector character of the monopole current, we should not expect
that the two models are in the same universality class, as pointed out in
\cite{KoKo92}).

To determine $\beta_{\rm perc}/\nu_{\rm perc}$ we investigated the scaling
with volume of $M_{\rm perc}$ at the (infinite volume) critical coupling.  The
results are very sensitive to the precise value of $\kappa_c$. We estimated
anyway $\beta_{\rm perc}/\nu_{\rm perc} =0.85(3)$ in good agreement with the
scaling relation
\begin{equation}
  \frac{\beta_{\rm perc}}{\nu_{\rm perc}} = \frac{1}{2}
  \left(d-\frac{\gamma_{\rm perc}}{\nu_{\rm perc}}\right)\,.
\end{equation}
The result for quenched non-compact QED is $\beta_{\rm perc}/\nu_{\rm perc} =
0.88(2)$.

\begin{figure}
  \begin{center}
    \epsfig{file=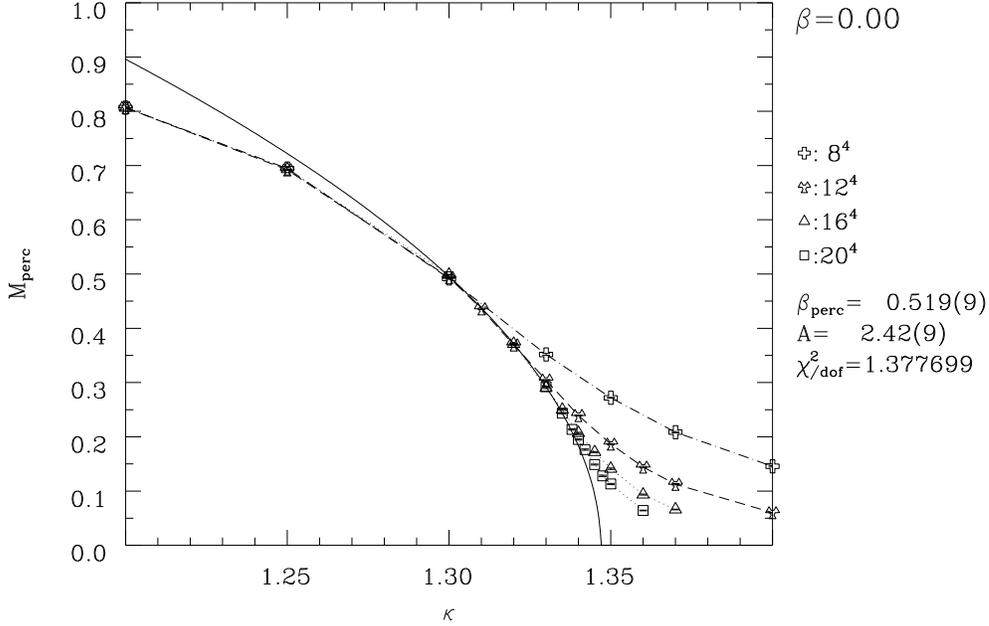,angle=90,width=\hsize,bbllx=65,bblly=15,bburx=540,bbury=780}%
    \caption{%
      Scaling behavior of $M_{\rm perc}$ for $\beta=0.00$. In the fit only
      data with $L\geq 16$ in the interval $\kappa=1.30 \ldots 1.33$ have been
      considered. The critical coupling was fixed to $\kappa_c=1.3469$.}
    \label{fig:larcon}
  \end{center}
\end{figure}

The exponent $\beta_{\rm perc}$ was determined from $M_{\rm perc}$ in the
phase with percolation, using the results not distorted by finite size
effects. The result for $\beta=0.00$ is shown in figure~\ref{fig:larcon}.
Here again we fixed the critical coupling. The fits give $\beta_{\rm
  perc}=0.52(2)$, where the error was determined by varying the critical
coupling in the interval $\kappa_c=1.346 \ldots 1.348$.  We repeated this
 for $\beta=0.30$  and obtained a
$\beta_{\rm perc}=0.50(4)$, smaller but compatible within errors.
($\beta = 0.58(2)$ in non-compact quenched QED 
\cite{KoKo92}.)%

Together with our determination of $\beta_{\rm perc}/\nu_{\rm perc} =0.85(3)$
we get the exponent for the correlation length $\nu_{\rm perc}=0.61(4)$ to be
contrasted with that of non-compact QED $\nu_{\rm perc} =0.66(3)$, and of pure
random site percolation $\nu_{\rm perc}=0.69(1)$.  A more careful analysis,
considering corrections to scaling, would be necessary to assess the
significance of these differences and to decide if percolation is in the same
universality class for all $\beta\leq 0.848$.  Differences, if any, are anyway
small.

\begin{figure}
  \begin{center}
    \epsfig{file=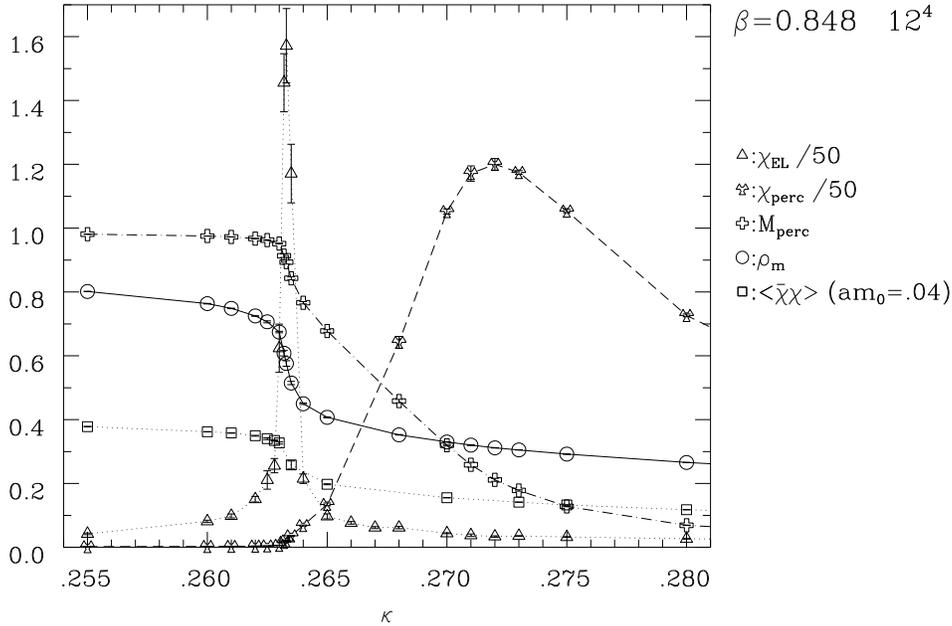,angle=90,width=\hsize,bbllx=65,bblly=15,bburx=540,bbury=780}%
    \caption{%
      Susceptibility of the monopole percolation $\chi_{\rm perc}$ and the
      link energy $\chi_{E_L}$, as well as the order parameter of the
      percolation $M_{\rm perc}$, the monopole density $\rho_m$ and the chiral
      condensate ${\langle\overline{\chi}\chi\rangle} (am_0=0.04)$ as function
      of $\kappa$ on a $12^4$ lattice at $\beta=0.848$ at the endpoint of the
      Higgs phase transitions.}
    \label{fig:perc}
  \end{center}
\end{figure}

Note that  for $\beta_{\rm E} = 0.848$ the percolation transition
at $\kappa=0.2714(3)$ is distinct from the Higgs phase transitions which
occurs at $\kappa_c=0.26333(1)$ \cite{AlAz93,Fr97a,FrJe97b}.  In
Fig.~\ref{fig:perc}, we see that all the observables are very sensitive to the
Higgs phase transition (note the sharp peak of the susceptibility of the link
energy), with the exception of the susceptibility  $\chi_{\rm perc}$
of the percolation. This shows that at the endpoint, the Higgs phase
transition and the percolation of the monopoles is uncorrelated. So the
percolation of the monopoles probably can not help us understand the
tricritical point in the dynamical model.%

\section{The chiral transition (NE) line}
\label{sec:chipt}

We consider first the behavior of the susceptibility ratio. The discussion
here aims at illustrating the general trend along the critical line.  In the
second subsection we cross check the results by fitting the data to the
equations of state. That includes a more detailed discussions on the selection
of the scaling window, and a careful error analysis.

\subsection{$R_\pi$}

We measure the ratio of $R_\pi$ as a function
of $am_0$: at the chiral transition $R_\pi$ should be
independent on $am_0$ if the transition is described by
a simple scaling law. Logarithmic corrections modify this behaviour. 

\begin{figure}\noindent
  \parbox{1cm}{(a)}\hspace{20mm}%
  \parbox[c]{10.5cm}{%
    \psfig{file=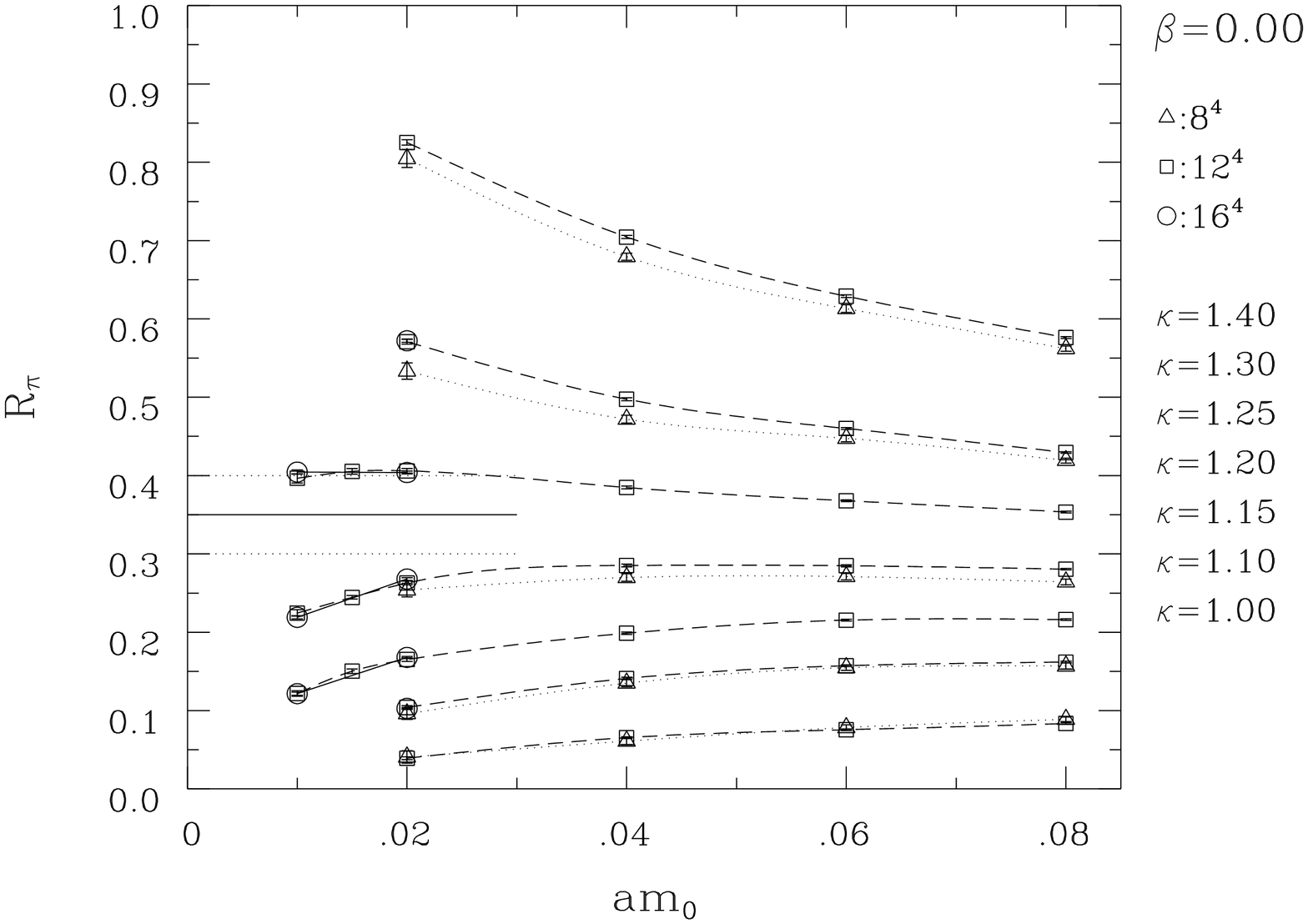,angle=90,width=10.5cm,%
      bbllx=65,bblly=15,bburx=540,bbury=780}}\\[2mm]%
  \parbox{1cm}{(b)}\hspace{20mm}%
  \parbox[c]{10.5cm}{%
    \psfig{file=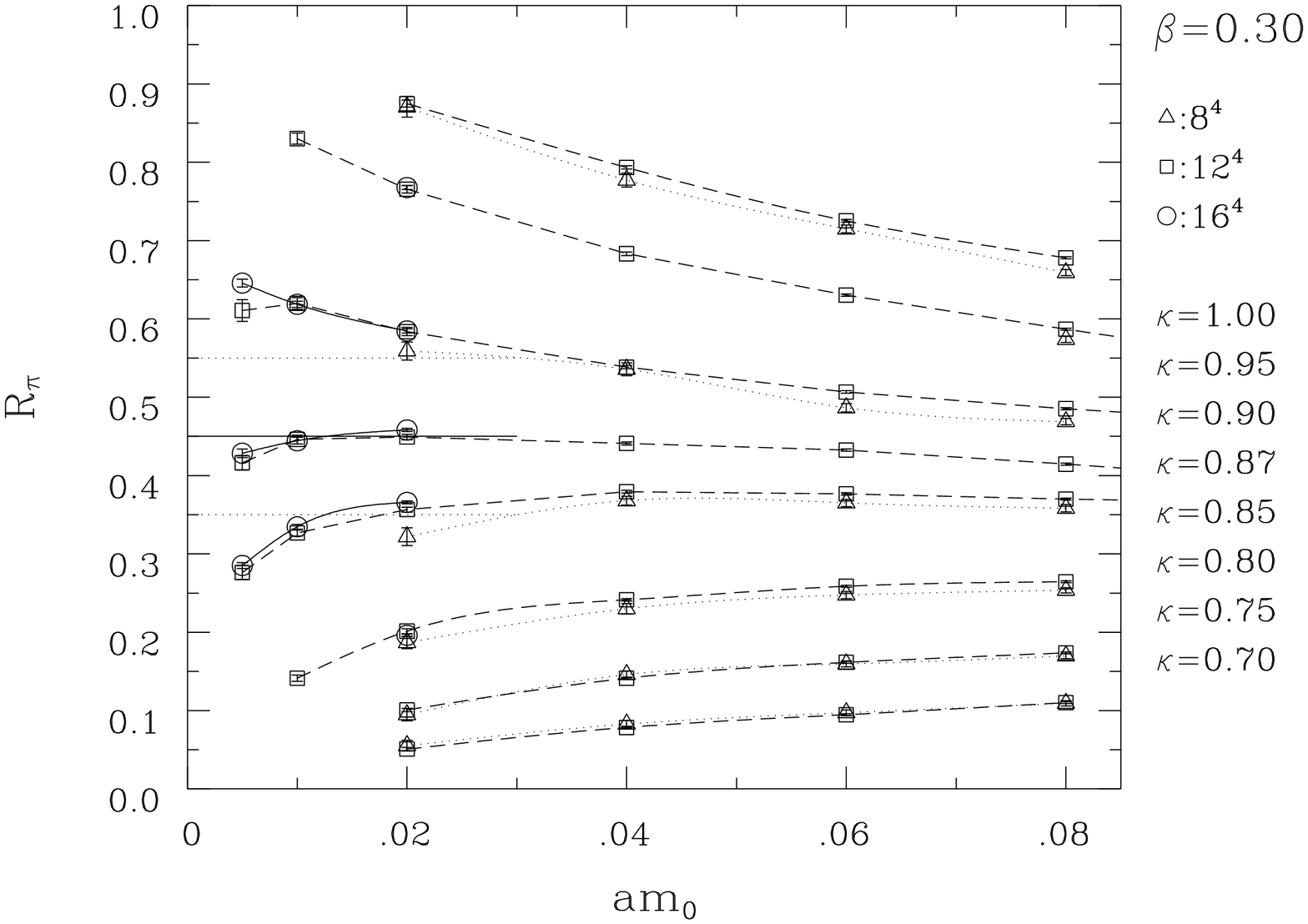,angle=90,width=10.5cm,%
      bbllx=65,bblly=15,bburx=540,bbury=780}}\\[2mm]%
  \parbox{1cm}{(c)}\hspace{20mm}%
  \parbox[c]{10.5cm}{%
    \psfig{file=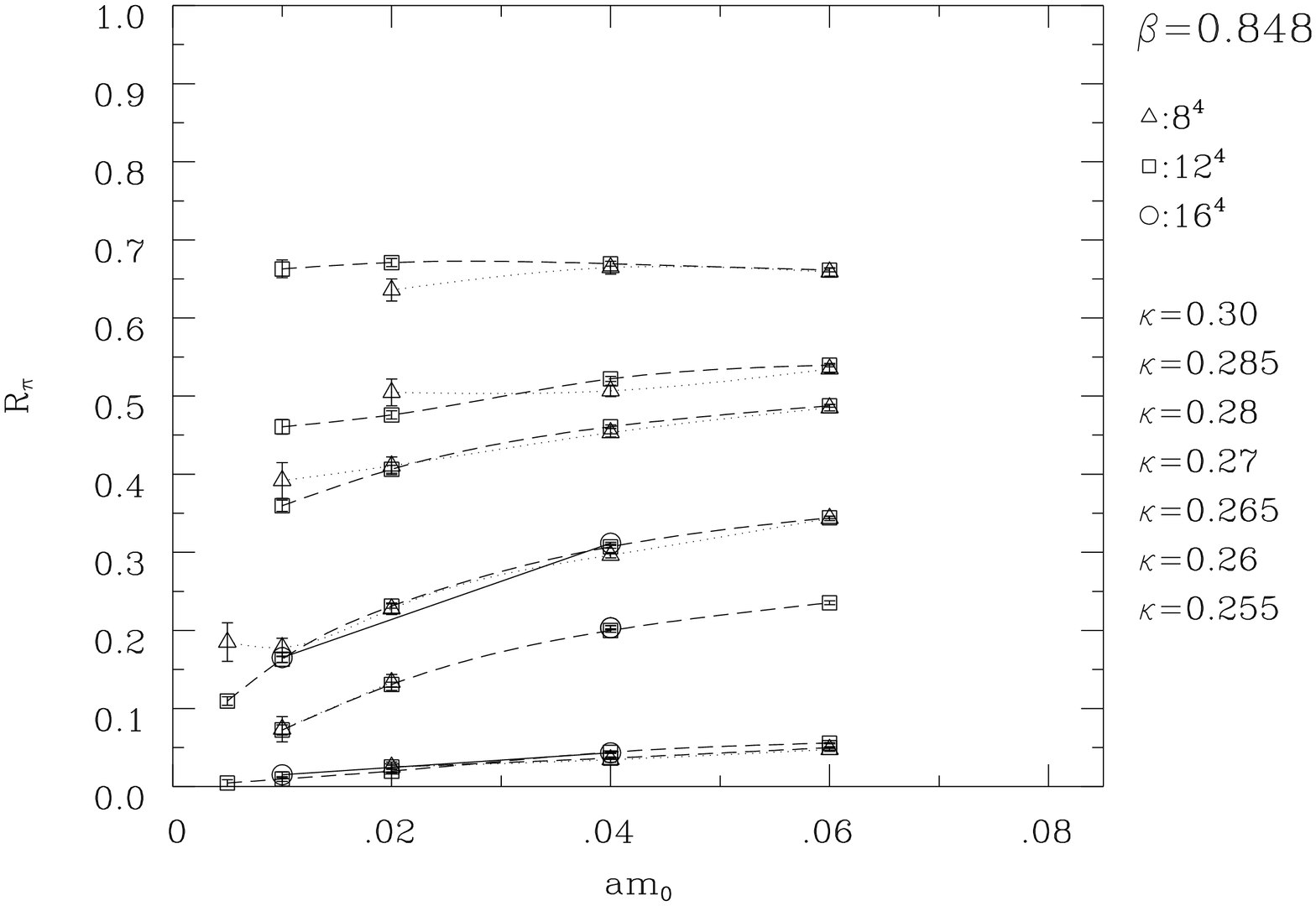,angle=90,width=10.5cm,%
      bbllx=65,bblly=15,bburx=540,bbury=780}}%
  \caption{%
    $R_\pi$ as function of $am_0$ at
    (a) $\beta=0.00$, (b) $\beta=0.30$ and (c) 
     $\beta=0.848\simeq\beta_E$. (Solid, dashed, dotted lines) are for
    ($16^4$, $12^4$, $8^4$) lattices. Approximate values for the critical
    point are indicated with horizontal lines.}
  \label{fig:qrpi}
\end{figure}%
For $\beta=0.00$ (Fig.~\ref{fig:qrpi}a) our data indicate a chiral transition
at $\kappa_c=1.23(3)$ with $R_\pi=0.35(5)$. This value compares well with the
expected mean field result $R_\pi(\kappa_c) = 1/\delta = 1/3$. Logarithmic
corrections, if any, are small, as we will see later on.

For $\beta=0.30$ (Fig.~\ref{fig:qrpi}b) the results suggest $\kappa_c=0.87(2)$
with $R_\pi(\kappa_c)=0.45(10)$. This corresponds to $\delta=2.2(6)$.
Logarithmic corrections \`a la Nambu--Jona Lasinio would predict
\begin{equation}
R_\pi (\kappa_c) = 1/ \big( 3 + 1 /\log(\cbcex)\big)
\end{equation}
resulting in a slight upward trend of the critical ratio, still compatible
with the data.  We comment more on this point in the next subsection.  The
curves for $am_0 > .02 - .04$ starts bending downwards, after the initial
rise: masses larger than .04 cannot be used safely to assess the critical
behaviour. In the next subsection we discuss in detail how to select the
appropriate scaling window.

For $\beta=0.848$ (point E) the Higgs phase transition is at $\kappa_c\simeq
0.263$.  We have shown above that the Higgs transition is separate from
monopole percolation, and here (Fig.~\ref{fig:qrpi}c) we see that at this
$\kappa$ a conventional second order chiral transition is incompatible with
the data.  The data in Fig.~\ref{fig:qrpi}c would rather suggest a chiral
transition in the interval [.28 -- .30], even on the right of the percolation
transition. But this behaviour is far from clear.

\begin{figure}
  \begin{center}
{\epsfig{file=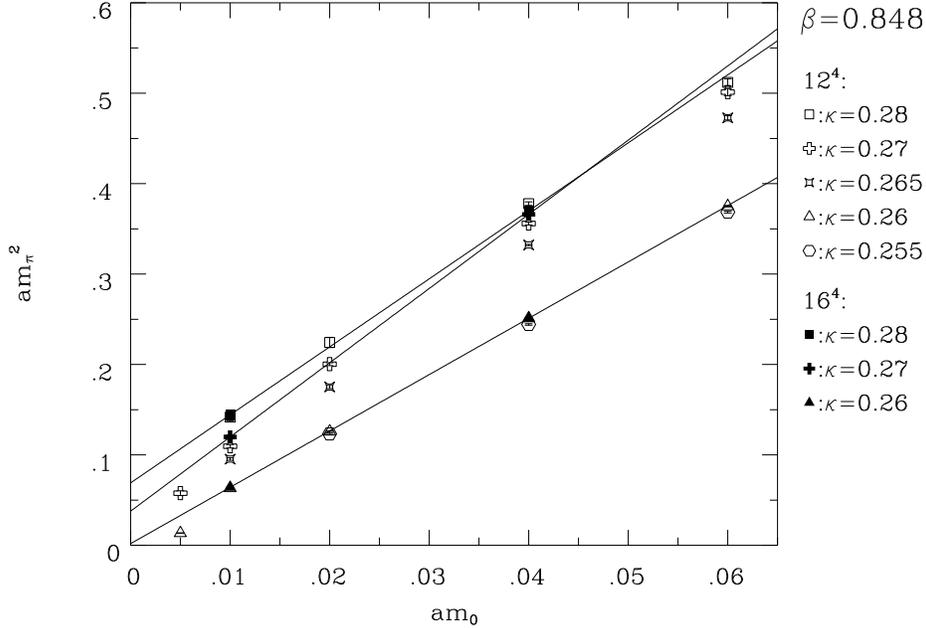,angle=90,width=\hsize,bbllx=65,bblly=15,bburx=540,bbury=780}}%
    \caption{%
      Behavior of the quadratic $\pi$ meson mass as function of $am_0$. In the
      broken phase a straight line through the origin is expected. The lines
      indicate a fit to the $16^4$ data.  For $\kappa=0.27$ we observed
      large finite size effects, which make the extrapolation unreliable.}
    \label{fig:qpit2}
  \end{center}
\end{figure}

As an alternative strategy to locate the chiral transition at $\beta = 0.848$
we investigated the scaling behavior of $(am_\pi)^2$ as a function of $am_0$
(Fig.~\ref{fig:qpit2}). For $\kappa=0.255$ and $\kappa=0.26$ the $\pi$ meson
mass scales corresponding to the PCAC relation $(am_\pi)^2\propto am_0$. For
larger $\kappa$, especally for those $\kappa$ between the transitions,
deviations from the PCAC relation can be observed.  The data for $\kappa=0.27$
on the $12^4$ lattice might suggest a straight line joining the origin and the
two smaller points. But the data for $am_0=0.01$ on the $16^4$ lattice
indicate that we have to be aware of finite size effects and the data on the
larger lattice favor a violated PCAC relation.  These ambiguities in the
extrapolations are not uncommon  in the numerical studies of phase
transition.  We conclude that also the PCAC analysis does not
reliably locate $\kappa_c$ of the chiral transition at $\beta = 0.848$.%

Because of the unusual behavior of the chiral condenstate the equation of
state also fails. So we cannot decide from this data if the Higgs and chiral
transition meet at the endpoint, as observed in the dynamical theory
\cite{FrFr95,FrJe97b}.

\subsection{Equation of State}

The results of the susceptibility ratio pose at least two questions: 1) is the
critical behaviour at $\beta = 0.30$ really different from the one at $\beta =
0.00$? 2) if yes, and the data at $\beta = 0.30$ are described by a power law
scaling, what are the other critical exponents? -- in particular, we are
interested in the relative position of percolation and chiral transition, and
in their correlation length exponents $\nu$.

The standard way to address these questions is to exploit the 
chiral equations of state (EOS). 
First we briefly discuss the results at $\beta
= 0.00$, which we shall use as a term of comparison, and then we will devote
more time to the discussion at $\beta= 0.30$.

The EOS framework is borrowed from ancient studies of
ferromagnetic transitions, and, as such, uses data in the
thermodynamic limit. Happily, this approach requires that we 
simulate the model within
its scaling window but not necessarily directly at the critical
point. Consequently it is possible to work with data free from finite
volume effects (which won't be possible {\it at} the critical 
point where a finite size scaling analysis would be mandatory).
Both the selection of the scaling window, and the control over finite
size effects are discussed below. 

Recall the possible critical behaviour, and the relative equations of state.
If scaling holds, the data are described by a `standard'
equation of state
\begin{equation}
am_0 = \cbcex^\delta f\big(t/\cbcex^{1/\beta_{\chi}}\big)\,,
\label{eq:scaling}
\end{equation}
 where $t$ is the reduced coupling (in our case $t=\kappa-\kappa_c$) and
whose first order approximation -- used in actual fits -- reads
\begin{equation}
\label{powerlaw}
am_0 = a^P (\kappa - \kappa_c) \cbcex^{\delta - 1/ \beta_{\chi}}+ b^P
\cbcex^\delta\,.
\label{eospow}
\end{equation}
 If the
theory is trivial, the most natural candidate to describe the data is an
equation of state \` a la Nambu-Jona-Lasinio.  
We will use the following form, which is motivated by the
leading term of the 1/N expansion \cite{KiKo94x}:
\begin{equation}
am_0 = a^{NJL} (\kappa - \kappa_c) \cbcex +  b^{NJL}\cbcex^3(\log \cbcex/s ) 
\label{eoslog}
\end{equation}
and we will just comment on its possible generalizations \cite{HaHa91,Zi91}
(in practical analysis $\log \cbcex/s$ will be replaced by $\log \cbcex + c$).

In many cases we used two different fitting procedures.  

For the first one we numerically invert eq.~(\ref{powerlaw}) for $\cbcex$ and
then make a minimum $\chi^2$ fit for the measured chiral condensate, taking
into account the errors.  We label the results obtained with this procedure
[F1].

The second one [F2] is a least squares procedure which minimizes $(am_0 -
am(\cbcex))^2$, $am(\cbcex))^2$ being the r.h.s.\ of either eq.~(\ref{eospow})
or eq.~(\ref{eoslog}).  We then compute the errors by jack-knifing the
results obtained by discarding  one point at a time. The
quality of the fit is estimated by 
$Q = (m_0 - m(\cbcex))^2/(\Delta(m(\cbcex)^2)N_{\rm points})$. 
For a good fit we expect $Q$  close to one.

\subsubsection{$\beta$ = 0.00}

As mentioned above, we expect that at $\beta = 0$ the model reduces
to a lattice NJL model. Recent numerical studies of four dimensional
NJL models  include \cite{KiKo94x,AlGo95,HaKo97}.

\begin{figure}
  \begin{center}
    \epsfig{file=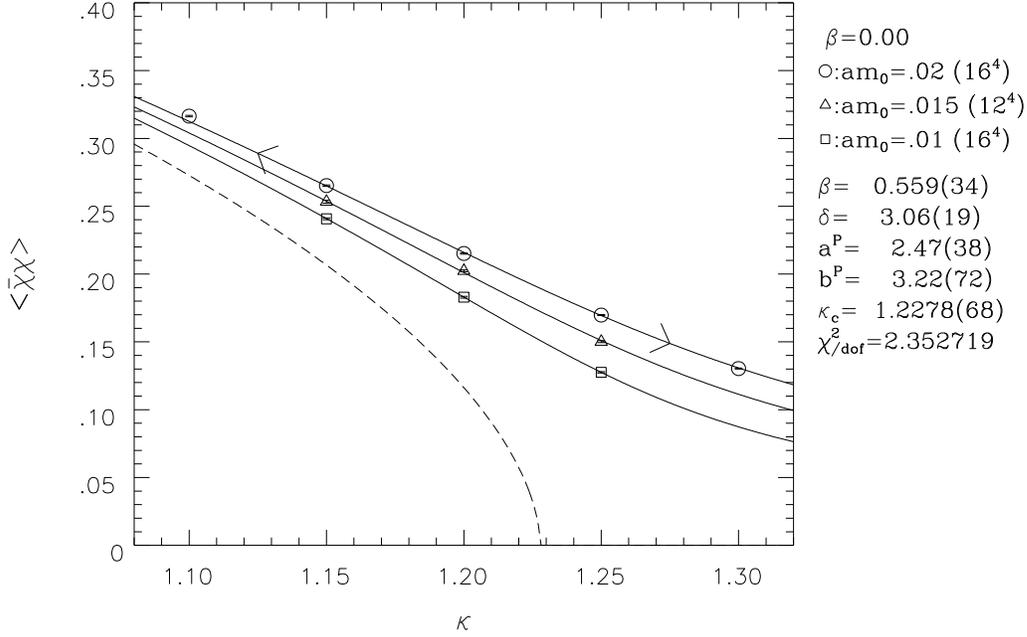,angle=90,width=\hsize,bbllx=65,bblly=15,bburx=540,bbury=780}
    \caption{Results of the power law fits described in the text at $\beta = 0.00$.
      The arrows indicate the fitted interval.}
    \label{fig:eosb000}
  \end{center}
\end{figure}

At $\beta = 0.00$ we restricted ourselves to $\kappa$ = (1.15, 1.20, 1.25).  A
power law fit with 5 free parameters which uses the data at bare masses .02
and .01 from the $16^4$ lattice, the data at .015 from the $12^4$ lattice is
shown in Figure \ref{fig:eosb000}.  The quality of the fit [F1] is nice, the
exponents $\delta = 3.06 (19)$, $\beta_\chi = 0.56(4)$ are all consistent with
mean field theory.  The critical coupling $\kappa_c$ is 1.228(7).%

We have then included logarithmic corrections, initially without scale (the
parameter $s=1$ in the EOS). We used data from the $12^4$ lattice, and the
procedure [F2]. In the same interval we obtain $\kappa_c = 1.259(4)$.  Adding
one further point at $\kappa = 1.30$, $am_0 = .02$ moves the central values to
$\kappa_c = 1.258$.  The inclusion of a free scale parameter in the fits moves
the critical point to $\kappa_c = 1.248$, closer to the power law results.  We
have checked that we can discard the points with bare mass .02 without
altering the results.  We can also constrain the exponents to their mean field
values without appreciably degrading the quality of the fits - that
corresponds to `cancelling'
the logs in the NJL equation of state.  $\kappa_c$
is then 1.224(8).

In conclusion, the data at $\beta = 0.00$ are well described, as expected, by
a mean field critical behaviour, with $\kappa_c$ in the range $1.22 - 1.26$,
in agreement with the ratio analysis.  This range of critical couplings 
is well separated from that of monopole percolation as can be seen from Table
\ref{tab:suspeak} and Fig.~\ref{fig:twodef0}.  Logarithmic corrections, if
any, are small, their inclusion improves the quality of the fits, but does not
alter $\kappa_c$ nor do they allow fits on a wider interval.  These critical
couplings are significantly different from those of the percolation
transition.

\subsubsection{$\beta$ = 0.30}

Our first task is to select a reliable data sample for our fits.  From the
results for $R_\pi$, we know that masses larger than $.02$ are too heavy.  On
the other hand, small masses, which require large lattices, are only available
for three $\kappa's$. All in all, our candidate window for fits narrows down
to the nine points at $\kappa = .85, .87, .90$ -- we occasionally checked the
stability of our results by including the two $am_0 = .02$ points at $\kappa =
.80$ and $.95$ but we did not venture any further.  By contrasting data in
this range obtained on a $8^4$, $12^4$, and a $16^4$ lattice, we checked that
finite volume effects are already small on a $8^4$ lattice, and that the
difference between results on a $12^4$ and on a $16^4$ lattice is
statistically not significant (the worst case is $am_0=0.005$ were the data
are separated by two standard deviations).  We used the data on the $16^4$
lattice for our fits.

\begin{figure}
  \begin{center}
    \epsfig{file=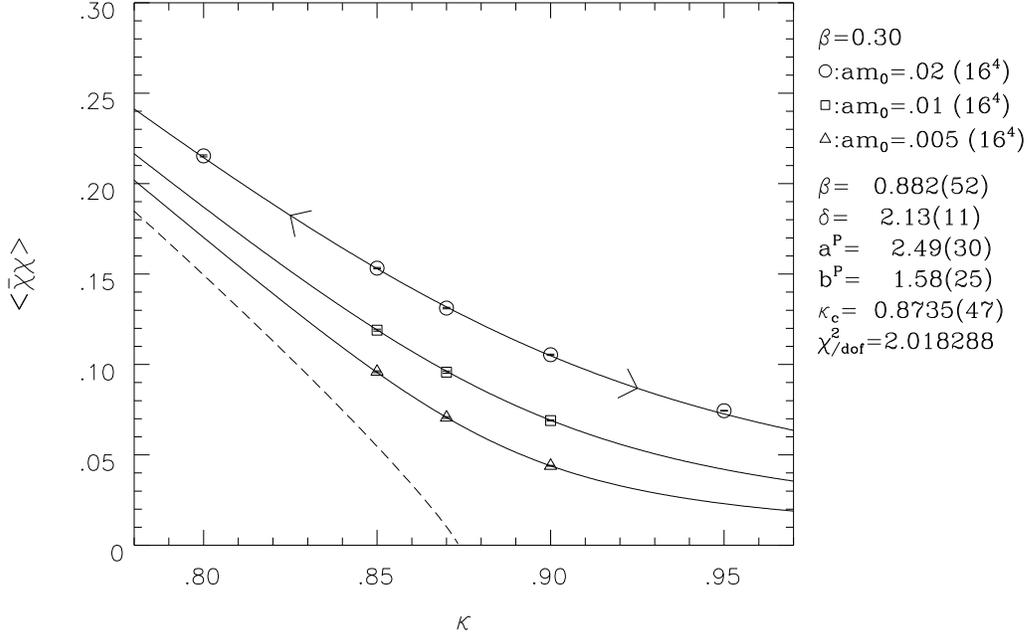,angle=90,width=\hsize,bbllx=65,bblly=15,bburx=540,bbury=780}
    \caption{As Figure \ref{fig:eosb000}, but $\beta = 0.30$. This fit
      correspond to the first line in table~\ref{tab:fitpow}.}
    \label{fig:eosb030}
  \end{center}
\end{figure}%
The results from [F1] applied to the power law form are shown in Figure
\ref{fig:eosb030}.  The stability of these results was checked by performing
the [F2] fits on the same range of parameters.  The results are shown in Table
\ref{tab:fitpow}, first line.  We get $\kappa_c=.874(6)$, $\delta=2.12(15)$,
and $\beta_\chi=0.88(5)$. (These exponents are very close to those of
quenched, non-compact QED: $\delta = 2.12(1)$, $\beta_\chi = 0.86(3)$, $\gamma
= 1.0$ \cite{KoKo93c,LoKo94}.)  Assuming scaling, these critical exponents
give $\nu = .695$.  We have verified that the fit is stable against the
inclusion of the two points at $am_0 = .02$, $\kappa = .80$ and .95.  Finally,
since we want to contrast the position of the chiral transition with the
position of the peaks of the monopole susceptibility, we have also tried fits
constraining the critical $\kappa = .882$, at the the lower bound of
$\kappa_c$ percolation. The results for the critical exponents (second line of
Table \ref{tab:fitpow}) are compatible with the previous ones and give $\nu =
.73$.  The quality parameter Q (1.79) is still acceptable.  Constraining
$\kappa_c$ deeper into the critical interval spanned by the critical $\kappa$
for percolation further degrades the quality of the fit, which remains
neverthless tolerable.  We have also searched for the best exponents assessing
by eye the quality of the scaling plots built following
eq.~(\ref{eq:scaling}). The best plot picks $\kappa_c = .868$, $\delta =
2.26$, $\beta_\chi = .8$, giving $\nu = .65$.%
\begin{table}
\caption{Results of the powerlaw EOS (eq.~(\ref{eospow})) fits at $\beta = 0.30$.}
\vspace{2mm}
\begin {tabular} {l l l l l | l } %\hline
$a^{P}$ & $b^{P}$   & $\kappa_c^{P}$ & $\delta$ & $\delta - 1/\beta_\chi$ & Q     \\ \hline
2.54(4) & 1.58(36)  & .874(6)       &2.12(15)   & .999(68)          & 1.068 \\  
2.19(28)& 1.22(7)   & .882[Fixed]   &1.94(2)    & .932(47)          & 1.79\\
2.00(40)& 1.08(8)   & .8875[Fixed]  &1.83(3)    & .893(82)          & 2.96 \\
%\hline
\end{tabular}
\label{tab:fitpow} 
\end{table}

We have also tried to fit the data with a logarithmic improved mean field
equation. The results of the fits are displayed in Table \ref{tab:fitlog}.
(Note, that this fit has   less free parameters.)  We see that the only
fit of quality comparable to the power law fits requires a variable scale. In
this case, the critical coupling is smaller than that from the power law fits:
$\kappa_c$ = .866(2) to be contrasted with $\kappa_c$ = .874 (6).  The shift
in the critical coupling with respect to the power law results can also be
appreciated by performing a constrained fit with $\kappa_c = .882$, which
fails completely (third line of Table \ref{tab:fitlog}, Q = 83.26): this
definitively shows that a NJL critical behaviour cannot be associated with the
percolation of magnetic monopoles.%
\begin{table}
\caption{Results of the logarithmic EOS (eq.~(\ref{eoslog})) fits at $\beta = 0.30$}
\vspace{2mm}
\begin {tabular} {l l l l | l } %\hline
$a^{NJL}$ & $b^{NJL}$   & $\kappa_c^{NJL}$ & $c$     & Q     \\ \hline 
 2.46(6)  &  -3.68(16)  &  .861(1)        &  0[Fixed]    & 6.94  \\
 2.54(5)  &  -6.12(84)  &  .866(2)        & .69(14)  & 1.18  \\
 2.81(19) & -13.88(2.01)&  .882[Fixed]    & 1.23(13) & 83.26    \\
%\hline
 \end{tabular}
\label{tab:fitlog} 
\end{table}

\begin{figure}
  \epsfig{file=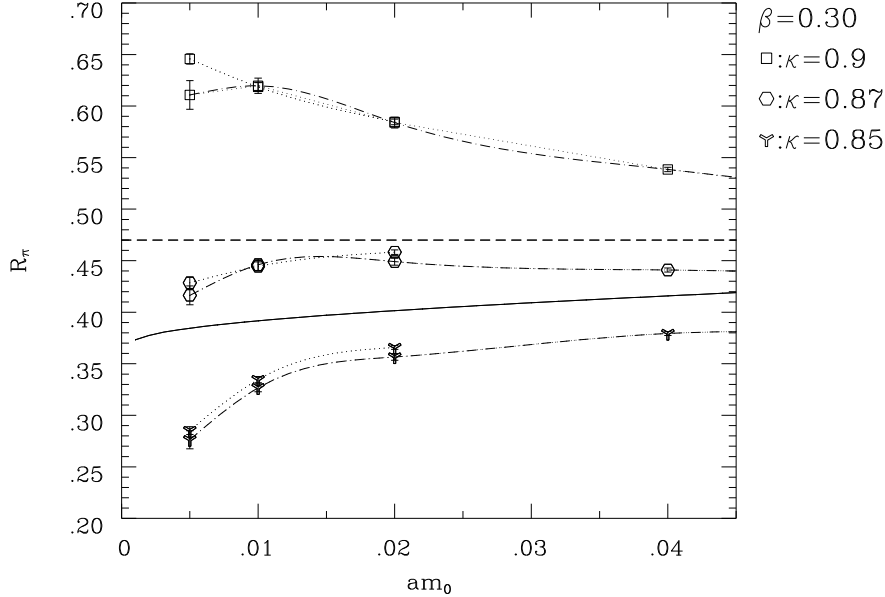,angle=90,width=.95\hsize,bbllx=65,bblly=15,bburx=540,bbury=780}
  \caption{$R_\pi$ in the critical region. Dotted is for the $16^4$
    lattice, dot--dash for the $12^4$. The dashed line is $R_\pi(\kappa_c) =
    1/\delta$ from the power law fit Table \ref{tab:fitpow}, the solid line is
    $R_\pi(\kappa_c)$ from the logarithmic fit, first line Table
    \ref{tab:fitlog}.}
  \label{fig:ratiob030}
\end{figure}

We can also study the sensitivity of the position of the critical point to the
equation of state by inspecting $R_\pi$. This gives a clearer indication
of the role played by the precise location of the critical point in determining the
critical scaling.  In Fig.\ \ref{fig:ratiob030} we show the results for
$R_\pi$ in the critical region constrasted with $1/\delta$ (solid line) from
our best power law fit, and with $R_\pi(\kappa_c)$ from the logarithmic fit
without scale. We see that the $R_\pi$'s corresponding to the two hypotheses
(logarithms and power law) fall in the $\kappa$ interval predicted by the
fits -- between .85 and .87 for log fits ($\kappa_c \simeq .86$), and between
.87 and .90 for power law ($\kappa_c \simeq .88$).%

\paragraph*{Summarizing:}
If the correct results (i.\,e.\ those describing the physics of the chiral
limit on infinite volumes) are given by the
logarithmic fits, second line Table \ref{tab:fitlog}, the percolation
transition and the chiral transition are distinct.  In this case we have not
learned anything about a possible role of magnetic monopoles in the existence
of strongly coupled theories.  Simply, the chiral transition would
be trivial and magnetic monopoles would have nothing to do with it, as it is
for $\beta=0.00$ and probably also for $\beta=0.848$.

If the correct results are given by power law fits, Table \ref{tab:fitpow},
the chiral transition coincides within errors with the percolation transition.
The critical exponent $\nu$ for the chiral transition is in the range .65 --
.73, to be contrasted with $\nu_{\rm perc}= .61(4)$, as we determined it in
section \ref{sec:perc}.  Taking into account the possibility of further
corrections, the exponents $\nu$ for the two transitions are compatible and
$\simeq 0.65$.

Note again these results differ from the ones in the full model where the
critical behavior along the line NE seems the same as the one of the trivial
NJL point. It is not in the scope of the present paper to investigate the
relationship between the quenched and full model, but this is certainly an
interesting subject for further investigations.

Finally, we might wonder about the sensitivity of the quality of the
logarithmic fits and/or the value of the critical coupling to further
corrections.
%One first consideration is that one may use a similar
%equation of state for the fermion mass rather than the chiral condensate
%\cite{KiKo94x}.  The corresponding exponents are then connected via scaling
%relations. Further more, subleading corrections may be different and should
%probably taken into account when considering improvements.  Unfortunately, the
%fermion mass shows much larger finite size effects than the chiral condensate
%and we failed to get reliable infinite volume estimates for it.
Typical corrections include powers of the logarithms. Although such
corrections are not there, up to O(1/$N^2$), they cannot be excluded in
general.  Our model, however, could be in the Yukawa universality class
\cite{HaHa91,Zi91}.  In this case, one can conceive of sizeable variation,
even in sign, of the exponents and the simplest parametrization of such
crossover behavior would suggest that the powers in the logs should be kept
free. It cannot be excluded that more general logarithmic equations of state
fit the data well, and certainly this would make the study of the
interrelation between the chiral transition and monopole percolation more
subtle.

\section{Summary}
We have investigated the topological excitations in the  phase
diagram of scalar QED. We confirmed with great precision the picture
suggested by \cite{RaKr83} and see clear signals for first order phase
transitions in the topological observables.

We have investigated in detail chiral transition and monopole percolation for
three points:

\begin{itemize}
\item $\beta = 0.00$: The chiral transition is described by mean field
  exponents, possibly with small (logarthmic) corrections.  The percolation
  transition is well separated from the chiral transition.
  
\item $\beta = 0.30$: Two possibilities are consistent with the data: 1$^{\rm
    st}$ a situation analogous to that at $\beta = 0.00$, but with large
  logarithmic corrections; 2$^{\rm nd}$ percolation and chiral transition
  coincident, with the same critical exponent $\nu$.

\item $\beta = 0.848$: The only clear statement here is that the Higgs
  transition and monopole percolation are distinct. We have not been able to
  measure with confidence the chiral transition, which has proven hard to
  study with conventional numerical methods.
\end{itemize}

Since the scalar field increases the order in the gauge fields and favours
chiral symmetry, separate chiral and monopole percolation transitions are not
unexpected: it might well be that monopoles percolate in the symmetric phase,
but their tendency to break chiral symmetry is overcome by the ordering
effects of the scalar fields.  This seems to be the case at $\beta = 0.00$,
and it is also possible at $\beta = 0.30$ (if the first possibility is
realized).  In this case chiral transition is most probably mean field like
along the whole line NE (without the point E), as it seems to be in the full
model.  Moreover, the chiral transition is separated from the percolation
transition: when this is the case monopoles should be irrelevant to the
dynamics of the chiral transition in the large volume, continuum limit.
Except for the point E (we reiterate that we don't have any conclusion on the
relative position of monopoles and chiral transition at the point E) this goes
hand in hand with the transition being trivial.  Still, the behaviour at
$\beta =0.30$ remains different from that at $\beta= 0.00$, and and it would
be interesting to find a coherent scenario accomodating these observations.

The scenario of \cite{KoWa96x} requires instead a chiral transition coincident
with the percolation transition, and sharing the same correlation length
exponent. When this occurs, the dynamics of the chiral transition should
inherit the characteristics of that of the magnetic monopoles, which survive
the non--trivial continuum limit. This could be the case at $\beta = 0.30$.
The scenario of \cite{KoWa96x} receives then some support from our
investigation.

Important questions concern the relevance of these results for the full model,
and, in general, the sensitivity to the number of flavours.  If the scenario
of \cite {KoWa96x} is realized, the situation is different from the one
observed in the full model: If this is the case, fermion screening plays an
important role in this model as it presumably does in the gauged Nambu Jona
Lasinio model: there the ladder (quenched) approximation predicts a non--mean
field scenario \cite{BaLo90,KoHa90}, while the full model is presumably
trivial\cite{KiKo97}.  These observations do not detract from the illustrative
value of our results at $\beta = 0.30$, but, of course, prevent us from
extending any of our conclusions to the full model at this stage of our
investigation.  If the full and quenched models really have different critical
behavior along the line NE, then we are challenged to understand something
unexpected and special about fermionic screening in this model. In the present
quenched model, the dynamical scalar field in the configurations completely
screens the electric charge, producing a vanishing renormalized gauge coupling
in this sector of the model.  If the fermions were also treated dynamically,
they would add nothing qualitatively new in terms of charge screening.
Therefore, the fermion--monopole interaction would have to have some essential
ingredient which renders the monopoles ineffective in driving chiral symmetry
breaking in the full model.  Simulations of non-compact QED with the number of
flavors $N_f$ varying from $2$ to $32$, do not show such an effect --- there
chiral symmetry breaking and monopole percolation were coincident, within
ample errorbars, for all $N_f$ \cite{KoWa96x}.  Better simulations of the full
and quenched models with smaller bare fermion masses and larger lattices are
needed now.

We should define better numerical strategies so to
show more clearly the correlation, or lack thereof, between chiral and
monopole observables.  It is however evident from our data, that for
$\beta=0.00$, when the chiral and monopole transitions are clearly distinct,
the chiral transition is clearly trivial and is not correlated with the
percolation of monopoles.  When the transitions grow closer for $\beta=0.30$,
so do the correlation length exponents $\nu$ (or the effective critical
exponent $\nu$ in the case of logarithmic triviality). The transitions may be
coincident, with the same critical exponent $\nu$.

\begin{ack}
  We thank J.~Jers\'ak for stimulating discussions, for reading the manuscript
  and for usuful comments.  This study used the codes developed for the
  \chupiv\ project for the field dynamics and the chiral observables, and the
  Hands--Wensley routine for the magnetic monopole meassurements.  Most of the
  computations have been performed on the CRAY-YMP and T90 of HLRZ J\"ulich,
  for some large lattices and small masses we used the CRAY-C90's of PSC and
  NERSC.  WF thanks HLRZ for hospitality. JBK thanks the Pittsburgh
  Supercomputer Center and the National Energy Research Supercomputer Center
  for access to their facilities.  MPL thanks the Physics Department of the
  University of Bielefeld for its hospitality. This work was supported by:
  DFG, the National Science Foundation, NSF-PHY9605199, the U.S.  Department
  of Energy (D.O.E.) under cooperative research agreement \#DF-FC02-94ER40818,
  and by Nato grant no.\ CRG 950896.
\end{ack}

\bibliographystyle{wunsnot}   % wunsnot style (unsorted numbers, no article titles)
%%\bibliographystyle{wunstit}   % wunstit style (unsorted numbers, article titles)
%%\bibliographystyle{wcitenot}  % wcitenot style (sorted cite$, no article titles)
%%\bibliographystyle{wcitetit}  % wcitetit style (sorted cite$, article titles)
%%\bibliographystyle{wabst}     % wabst style (unsorted cite$, article titles, abstracts, ...)
% \bibliography{jourabbr,ahpaper,our-papers,u1,referen,gauge,yukawa}
% begin_of_bibliography

% end_of_bibliography

\end{document}